\documentclass[a4paper,fleqn]{cas-dc}

\usepackage[numbers]{natbib}

\usepackage{caption}

\def\tsc#1{\csdef{#1}{\textsc{\lowercase{#1}}\xspace}}
\tsc{WGM}
\tsc{QE}
\tsc{EP}
\tsc{PMS}
\tsc{BEC}
\tsc{DE}

\begin{document}
\let\WriteBookmarks\relax
\def\floatpagepagefraction{1}
\def\textpagefraction{.001}

\shorttitle{SOC}

\shortauthors{Salva\~na \& Tangonan (2025)}

\title [mode = title]{Predicting Power Grid Failures Using Self-Organized Criticality: A Case Study of the Texas Grid 2014-2022}                   

%
\author[1,2]{Mary Lai O. Salva\~na}[orcid=0000-0003-4868-7713]

\cormark[1]

\ead{marylai.salvana@uconn.edu}

\ead[url]{marylaisalvana.com}

\address[1]{Department of Statistics, University of Connecticut, 215 Glenbrook Rd., Storrs, Connecticut, 06269, USA}

\author[2]{Gregory L. Tangonan}
\ead{goriot@mac.com}

\address[2]{Ateneo Innovation Center, Ateneo de Manila University, Quezon City, Metro Manila, 1108, Philippines}

\cortext[cor1]{Corresponding author}

\begin{abstract}
This study develops a novel predictive framework for power grid vulnerability based on the statistical signatures of Self-Organized Criticality (SOC). By analyzing the evolution of the power law critical exponents ($\alpha$) in outage size distributions from the Texas grid during 2014-2022, we demonstrate the method's ability for forecasting system-wide vulnerability to catastrophic failures. Our results reveal a systematic decline in the critical exponent from 1.45 in 2018 to 0.95 in 2020, followed by a drop below the theoretical critical threshold ($\alpha$ = 1) to 0.62 in 2021, coinciding precisely with the catastrophic February 2021 power crisis. Such predictive signal emerged 6-12 months before the crisis. By monitoring critical exponent transitions through subcritical and supercritical regimes, we provide quantitative early warning capabilities for catastrophic infrastructure failures, with significant implications for grid resilience planning, risk assessment, and emergency preparedness in increasingly stressed power systems.
\end{abstract}



\maketitle

\section{Introduction}

Power grid failures represent an increasingly critical socioeconomic risk as climate change intensifies extreme weather events and infrastructure dependencies deepen. The February 2021 Texas power crisis—which resulted in over 200 deaths and approximately \$130 billion in economic damage—highlighted a critical need for improved predictive capabilities in identifying system-wide vulnerability before catastrophic failures occur \citep{txc_winterstorm_2021}.
Traditional predictive approaches to grid reliability typically include fragility models \citep{karagiannakis2025fragility}, spatio-temporal correlation analysis \citep{lee2023quantifying, bhusal2024estimating}, machine learning and deep learning methods \citep{xie2020review, madasthu2023ensemble, yang2023assessing, wang2025deep}. However, these approaches failed to anticipate the catastrophic failure of the Texas power grid in 2021 due to several fundamental limitations. Fragility models generally focus on individual component resilience rather than emergent system-wide behaviors. Correlation-based approaches often assume historical patterns will persist, making them inadequate when faced with unprecedented conditions. Even sophisticated machine learning techniques struggle with the "black swan" nature of such events, as they rely heavily on training data that may not capture the complex interactions and cascading effects that manifest during extreme scenarios.

The inherent difficulty in predicting low-probability, high-consequence events stems from their statistical rarity and the absence of comprehensive historical records. Power systems increasingly operate in regimes for which no precedent exists, driven by changing climate patterns, evolving energy mixes, and shifting demand profiles. The Texas grid failure exemplified this challenge—traditional reliability metrics failed to signal the impending system-wide collapse \citep{busby2021cascading}. Furthermore, traditional methods typically assess risk through deterministic or probabilistic frameworks that struggle to capture how small, seemingly manageable disturbances can trigger disproportionately large cascading failures when systems approach critical states.

This fundamental gap in predictive capability necessitates novel approaches that can detect early warning signals of catastrophic failure in complex systems. One promising direction draws from complexity science, particularly the theory of Self-Organized Criticality (SOC), which offers a framework for understanding how complex systems naturally evolve toward critical states characterized by power-law distributions of event sizes \citep{bak1996nature, dobson2007complex}. By analyzing the statistical signatures of SOC in historical outage data, we can potentially identify when power grids approach dangerous thresholds before catastrophic failures materialize \citep{carreras2004evidence}.

This paper develops a predictive framework based on SOC theory, which proposes that complex systems with many interacting components naturally evolve toward critical states where small perturbations can trigger large-scale events. 
We demonstrate that this approach successfully predicted the conditions leading to the 2021 Texas power crisis, providing a 6-12 month advance warning that the system was approaching a critical threshold where catastrophic failures become exponentially more likely. Beyond theoretical contributions, our predictive methodology offers practical applications for grid operators, emergency managers, and financial risk assessments. By quantifying grid vulnerability through critical exponent monitoring, this approach enables proactive interventions before catastrophic failures occur, potentially saving lives and billions in economic damages.

\section{Methodology}

\subsection{The SOC Theory}

First proposed by \cite{bak1996nature}, SOC theory describes how complex systems with many interacting components naturally evolve toward critical states—poised between order and disorder—where small perturbations can trigger large-scale cascading events. At criticality, the probability distribution of event sizes follows a power law:
\begin{equation}
P(x) \propto x^{-\alpha}
\end{equation}
where $x$ represents event size (in our case, number of customers affected by an outage) and $\alpha$ is the critical exponent. This power-law signature has been observed in numerous complex systems, including earthquakes, forest fires, financial markets, and power grids \citep{sornette2006critical, dobson2007complex}.

The critical exponent $\alpha$ serves as a key parameter characterizing system dynamics. When $\alpha > 1$, the power-law distribution is normalizable and the system exhibits well-defined probabilistic behavior with statistically rare extreme events. However, as $\alpha$ approaches and falls below 1, the distribution becomes increasingly heavy-tailed, and the relative probability of extreme events grows without bound \citep{pruessner2012self}. This property makes $\alpha$ a potential leading indicator for critical transitions in complex systems.

Previous work by \cite{carreras2004evidence} and \cite{dobson2007complex} identified power-law distributions in historical blackout data, suggesting that power grids may indeed operate as self-organized critical systems. Our approach extends this work by developing a temporal analysis framework to track the evolution of $\alpha$ and evaluate its predictive capability for catastrophic events.

\subsection{Data Collection and Processing}

We retrieve outage data from the publicly available Oak Ridge National Laboratory's Environment for Analysis of Geo-Located Energy Information (EAGLE-I) power outage dataset \citep{brelsford2024dataset}. EAGLE-I has been collecting U.S. electric outage data since 2014, providing a comprehensive historical record of power disruptions across USA. The dataset includes:
\begin{itemize}
\item Outage counts representing the number of affected customers at the county level
\item Temporal resolution with data snapshots recorded every 15 minutes
\item Geographic information down to the county level
\item Utility provider information for each service area
\end{itemize}
For this study, we focused exclusively on outage data from Texas covering the period from January 2014 to December 2022. This temporal range encompasses the catastrophic February 2021 Texas power crisis, providing an opportunity to analyze system behavior before, during, and after this significant event.

For the extreme events analysis, we used daily aggregation, selecting only the maximum outage event for each day to avoid double-counting related outages and to focus on the most significant disruptions. This approach aligns with the focus of SOC theory on event size distributions rather than the precise timing of individual outages.

\subsection{Power-Law Parameter Estimation}
To estimate the critical exponent $\alpha$, we employed log-linear regression on the frequency distribution of outage sizes using backward-looking temporal windows. For each time point $t$, we analyzed data from the preceding 180-day window (from $t-180$ days to $t$), performing the following steps:
\begin{enumerate}
\item Filtered the Texas outage data for the specific time window.
\item Constructed a frequency distribution by counting the occurrences of each distinct outage size (customers affected)
\item Applied log-transformation to both the outage size and frequency variables
\item Fitted a linear regression model to the log-transformed data using:
\end{enumerate}
\begin{equation}
\ln(f_i) = C - \alpha \ln(x_i) + \epsilon_i
\end{equation}
where $f_i$ is the frequency of outage size $x_i$, $C$ is a constant, $\alpha$ is the critical exponent (the negative of the slope coefficient), and $\epsilon_i$ represents the error term.
The quality of fit was assessed using the coefficient of determination ($R^2$), which quantifies the proportion of variance in the log-frequency explained by the log-transformed outage size.
We conducted the analysis with both 180-day and 90-day backward-looking windows (using 30-day and 15-day step sizes, respectively) to examine the sensitivity of our results to the choice of temporal scale.

\subsection{Temporal Evolution of System Criticality}

The temporal evolution of power system vulnerability can be characterized through systematic tracking of the critical exponent, $\alpha$. Building on theoretical foundations in complexity science \citep{bak1996nature, dobson2007complex}, we developed a methodology to quantify how system criticality evolves over time:

\begin{enumerate}
\item For each time point $t$, we estimated the critical exponent $\alpha_t$ from the power-law distribution of outage sizes in the preceding window (from $t-\Delta t$ to $t$)
\item Each $\alpha_t$ value thus represents the system state at time $t$ based on the preceding 180 days of outage data
\item This backward-looking approach ensures that each critical exponent characterizes the system's state using only historical information available at that point in time
\end{enumerate}
This approach allowed us to identify critical slowing down and critical fluctuations—hallmark signatures of systems approaching phase transitions \citep{scheffer2009early}. In the context of power grid reliability, a phase transition represents the shift from a regime where outages remain localized and manageable to one where they propagate throughout the system in cascading failures with potentially catastrophic consequences. By tracking the critical exponent's trajectory relative to the theoretical threshold ($\alpha = 1$), we could observe the system's progression toward criticality, characterized by power-law distributions with increasingly heavier tails.

\subsection{Critical Transition Detection}

Complex systems, including power grids, can experience abrupt shifts from one stable state to another, known as critical transitions \citep{scheffer2009early, dakos2015resilience}. The core challenge in complexity science is identifying early warning signals before these transitions occur. Our approach addresses this challenge by analyzing the critical exponent $\alpha$ derived from the power-law distribution of outage sizes.

In SOC theory, the critical exponent $\alpha$ determines the probability distribution of event magnitudes. When $\alpha$ is large, the probability of large-scale events drops rapidly, making them extremely rare. Conversely, when $\alpha$ is small, the decline in probability is more gradual, making larger events relatively more probable. 

The value $\alpha = 1$ represents a theoretically significant threshold. For a power-law distribution $P(x) \propto x^{-\alpha}$, the normalization condition requires:
\begin{equation}
\int_{x_{min}}^{\infty} P(x) \, dx \propto \int_{x_{min}}^{\infty} x^{-\alpha} \, dx = \frac{x^{1-\alpha}}{1-\alpha} \Big|_{x_{min}}^{\infty}
\end{equation}
This integral converges only when $\alpha > 1$. For $\alpha \leq 1$, the integral diverges, meaning the probability mass becomes unbounded toward large events. In practical terms, when $\alpha$ falls below 1, the system enters a regime where extreme events are no longer statistically ``rare''---they increasingly dominate the distribution and become inevitable as observation windows extend.

Our framework for detecting critical transitions focuses on three key aspects of the critical exponent's dynamics:

\begin{itemize}
    \item Crossings of the $\alpha = 1$ threshold, signaling transitions into regimes of heightened vulnerability
    \item The duration of periods when the system remains below this threshold
    \item The depth of excursion below the threshold (how far $\alpha$ drops below 1)
\end{itemize}

We hypothesize that these three aspects---threshold crossings, persistence, and excursion depth---together characterize the vulnerability of power systems to catastrophic failures. This framework provides a foundation for interpreting the critical exponent's behavior in relation to system-wide risk and offers a quantitative approach to identifying periods of heightened vulnerability before catastrophic failures materialize.

The February 2021 Texas power crisis provides a unique case study for testing these predictions. By analyzing how the critical exponent evolved relative to the $\alpha = 1$ threshold before, during, and after this well-documented catastrophic failure, we can evaluate whether threshold monitoring provides actionable early warning signals for cascading failures in power systems.

\section{Results}

\begin{figure}
    \centering
    \begin{minipage}[b]{0.45\textwidth}
        \centering
        \includegraphics[width=\textwidth]{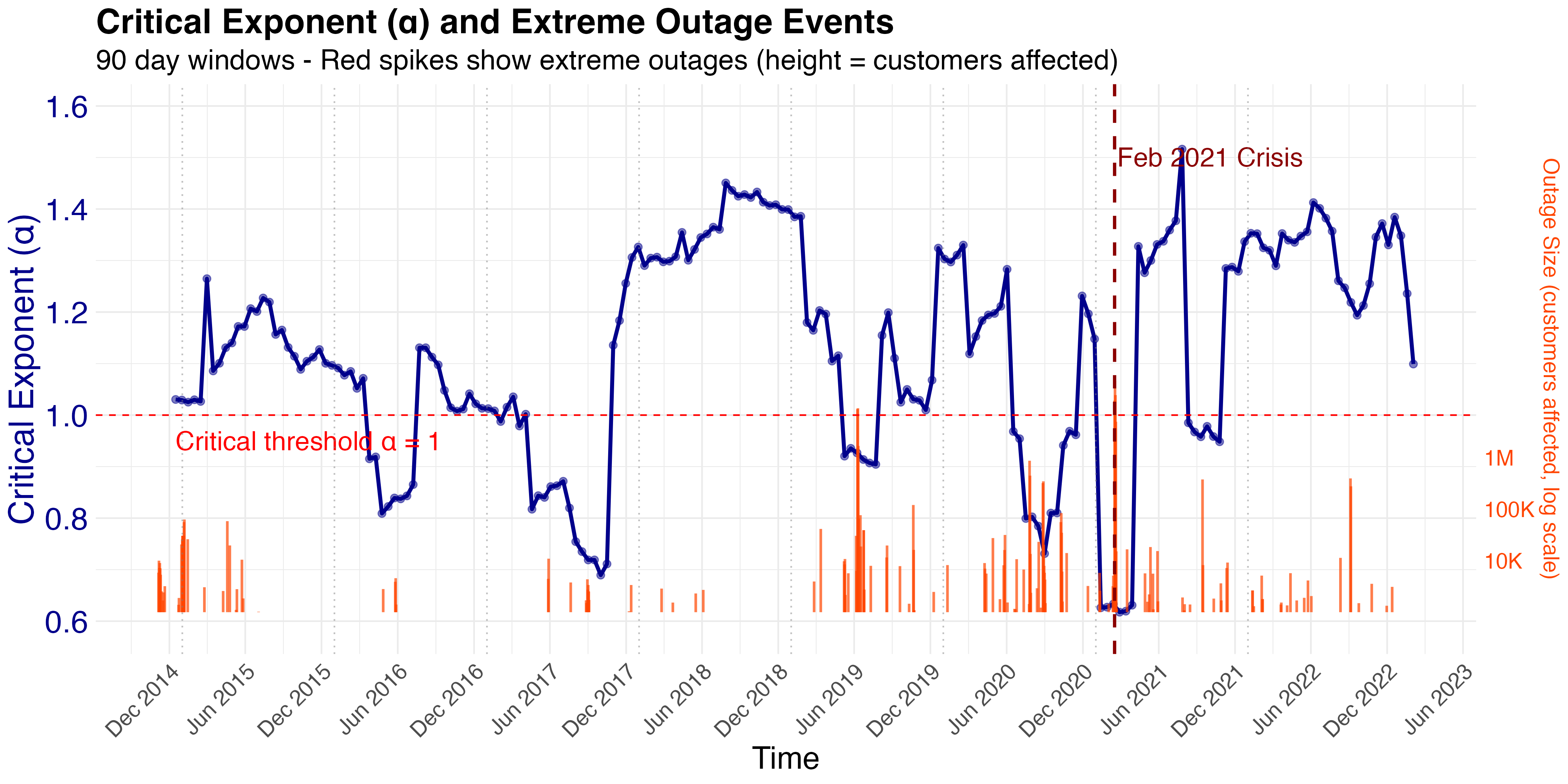}
        \text{(a) 90-day window}
        \label{fig:alpha_90d}
    \end{minipage}
    \hfill
    \begin{minipage}[b]{0.45\textwidth}
        \centering
        \includegraphics[width=\textwidth]{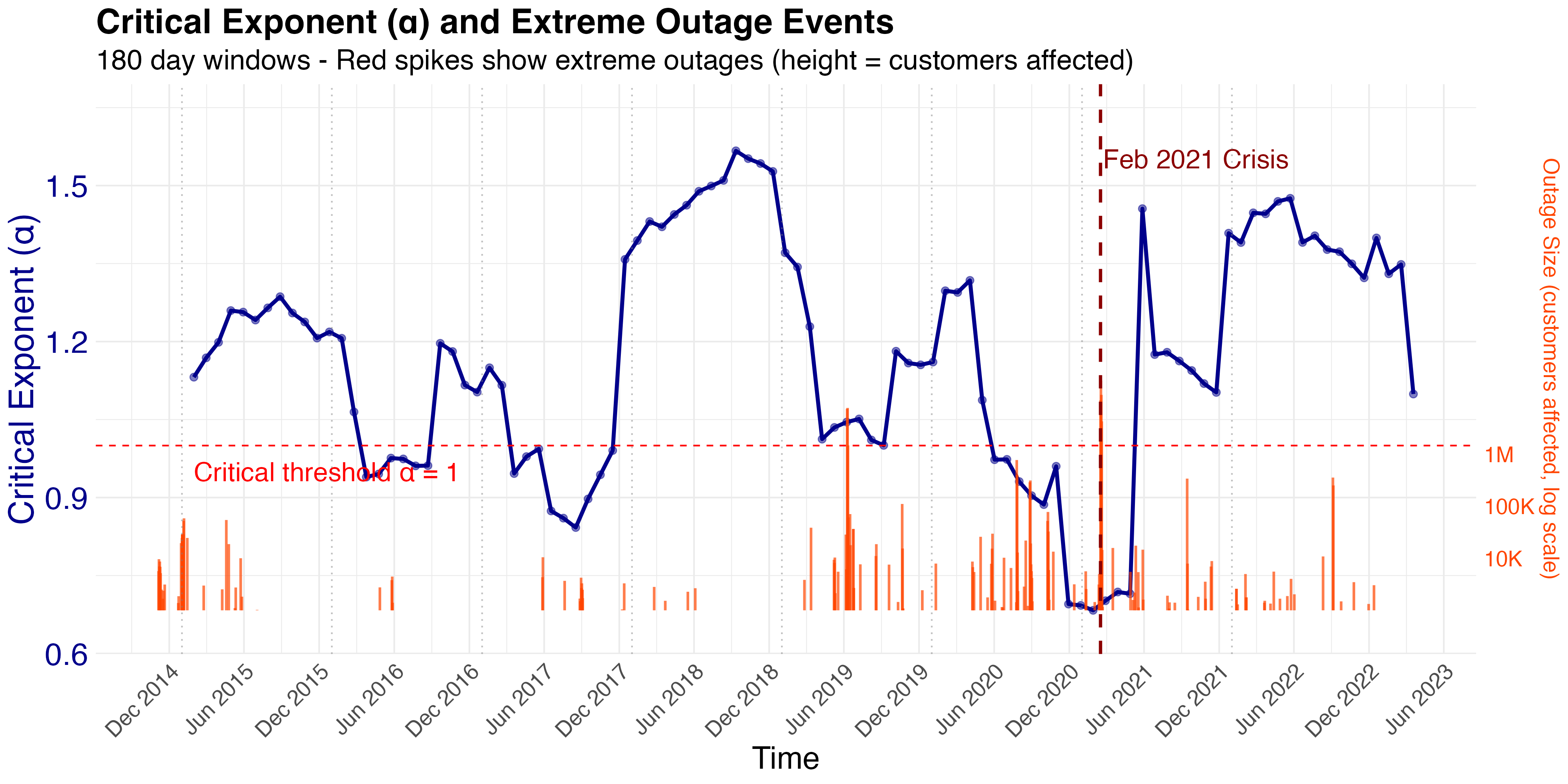}
        \text{(b) 180-day window}
        \label{fig:alpha_180d}
    \end{minipage}
    \caption{Temporal evolution of the critical exponent ($\alpha$) for the Texas power grid from 2014-2023. Red vertical spikes represent extreme outage events (99.9th percentile), with heights proportional to the logarithm of customers affected. The horizontal red dashed line indicates the critical threshold $\alpha = 1$, below which the system becomes increasingly vulnerable to catastrophic failures. The vertical red dashed line marks the February 2021 power crisis. Both window sizes reveal a significant drop in $\alpha$ below the critical threshold in the months preceding the crisis, providing an early warning signal of impending system-wide failure.}
    \label{fig:alpha_comparison}
\end{figure}

\subsection{Temporal Evolution of Critical Exponents}

The analysis of Texas power grid outage data from 2014-2022 reveals distinct patterns in the critical exponent $\alpha$ that align with periods of system vulnerability. Figure~\ref{fig:alpha_comparison} displays the temporal evolution of the critical exponent using 90-day and 180-day windows, respectively, alongside extreme outage events (represented by vertical red spikes).

Both window sizes demonstrate similar overall patterns, with the 180-day window providing smoother transitions that highlight longer-term vulnerability trends, while the 90-day window captures more rapid fluctuations in the system state. Several key observations emerge from these results:

\begin{enumerate}
    \item The critical exponent $\alpha$ exhibits substantial temporal variation, ranging from approximately 0.6 to 1.6 over the study period.
    
    \item Multiple periods where $\alpha$ drops below the critical threshold of 1 are observed throughout the dataset, indicating recurrent intervals of increased system vulnerability.
    
    \item The period immediately preceding the February 2021 Texas power crisis shows a pronounced decline in $\alpha$, reaching its lowest value of approximately 0.62—significantly below the critical threshold of 1.
    
    \item A systematic decline in the critical exponent from approximately 1.3 in early 2019 to below 1.0 by mid-2020 is evident, suggesting a gradual deterioration of system resilience in the months preceding the catastrophic failure.
    
    \item Following the February 2021 crisis, the critical exponent rapidly increased, suggesting a system-wide reconfiguration or adaptation that temporarily improved resilience before returning to fluctuations above and below the critical threshold.
\end{enumerate}

\subsection{Critical Threshold Crossings and Extreme Events}

Of particular significance is the relationship between critical threshold crossings ($\alpha < 1$) and subsequent extreme outage events. Our analysis reveals a lead time of approximately 6-9 months between the critical exponent's drop below 1 in mid-2020 and the catastrophic February 2021 failure. This pattern suggests the utility of the critical exponent as an early warning indicator of system-wide vulnerability.

When comparing the 90-day and 180-day window analyses:

\begin{enumerate}
    \item The 90-day window shows greater sensitivity to short-term fluctuations, with more frequent threshold crossings that might indicate heightened short-term risk.
    
    \item The 180-day window provides a more stable signal with clearer indications of persistent vulnerability, particularly evident in the sustained decline preceding the February 2021 crisis.
    
    \item Both analyses captured the critical signal preceding the 2021 crisis, with $\alpha$ dropping well below 1 in the months leading up to the event.
\end{enumerate}

\subsection{Extreme Event Analysis}

We identified extreme outage events as those exceeding the 99.9th percentile of customers affected, representing the most severe disruptions in the Texas power grid over the study period. These events, shown as vertical red spikes in Figure~\ref{fig:alpha_comparison}, demonstrate a notable pattern:

\begin{enumerate}
    \item Extreme events occur with higher frequency and magnitude during periods when the critical exponent is below or approaching the critical threshold of $\alpha = 1$.
    
    \item The 2014-2015 period shows clusters of extreme events when $\alpha$ was fluctuating near the critical threshold.
    
    \item A relative stability period in 2017-2018 corresponds with consistently higher $\alpha$ values (approximately 1.3-1.6) and fewer extreme events.
    
    \item The increasing frequency of extreme events in 2019-2020 coincides with the systematic decline in $\alpha$ leading up to the February 2021 crisis.
    
    \item The February 2021 crisis itself represents the most extreme event in the dataset, affecting approximately 4.5 million customers at its peak.
\end{enumerate}

This pattern demonstrates that extreme events are not simply random occurrences but are more likely to happen when the system's critical exponent indicates heightened vulnerability. The logarithmic scale of the outage size axis (right vertical axis in Figure~\ref{fig:alpha_comparison}) highlights that these extreme events span several orders of magnitude, from approximately 10,000 to over 1 million customers affected.

The clustering of extreme events around periods where $\alpha < 1$ provides empirical support for the theoretical prediction that systems with power-law exponents below 1 are prone to catastrophic, system-spanning events. This relationship offers a potential mechanism for predicting not just the likelihood but also the potential severity of future power grid failures.

\section{Conclusion}

This study establishes a novel predictive framework for power grid vulnerability based on the statistical signatures of Self-Organized Criticality (SOC). By tracking the evolution of power law critical exponents ($\alpha$) in outage size distributions from the Texas grid, we have demonstrated an effective methodology for forecasting system-wide vulnerability to catastrophic failures.

The implications of this research extend beyond the specific case of the Texas power grid. Our methodology can be applied to other power systems and potentially to different types of critical infrastructure networks that exhibit characteristics of self-organized criticality. By monitoring critical exponents derived from outage distributions, operators and policymakers can gain valuable lead time to implement preventive measures before catastrophic failures occur.

Future research should expand this approach by incorporating additional variables such as weather patterns and load profiles, modeling the power grid network according to SOC principles, developing real-time monitoring systems, and identifying intervention strategies that can shift power systems away from vulnerable states when warning signals appear. By identifying statistical precursors to catastrophic failures, this framework bridges the gap between theoretical complexity science and practical infrastructure management, contributing to more resilient power grids in an era of increasing climate uncertainties.

\bibliographystyle{IEEEtran}

\bibliography{cas-refs}

\vspace{12cm}

\bio{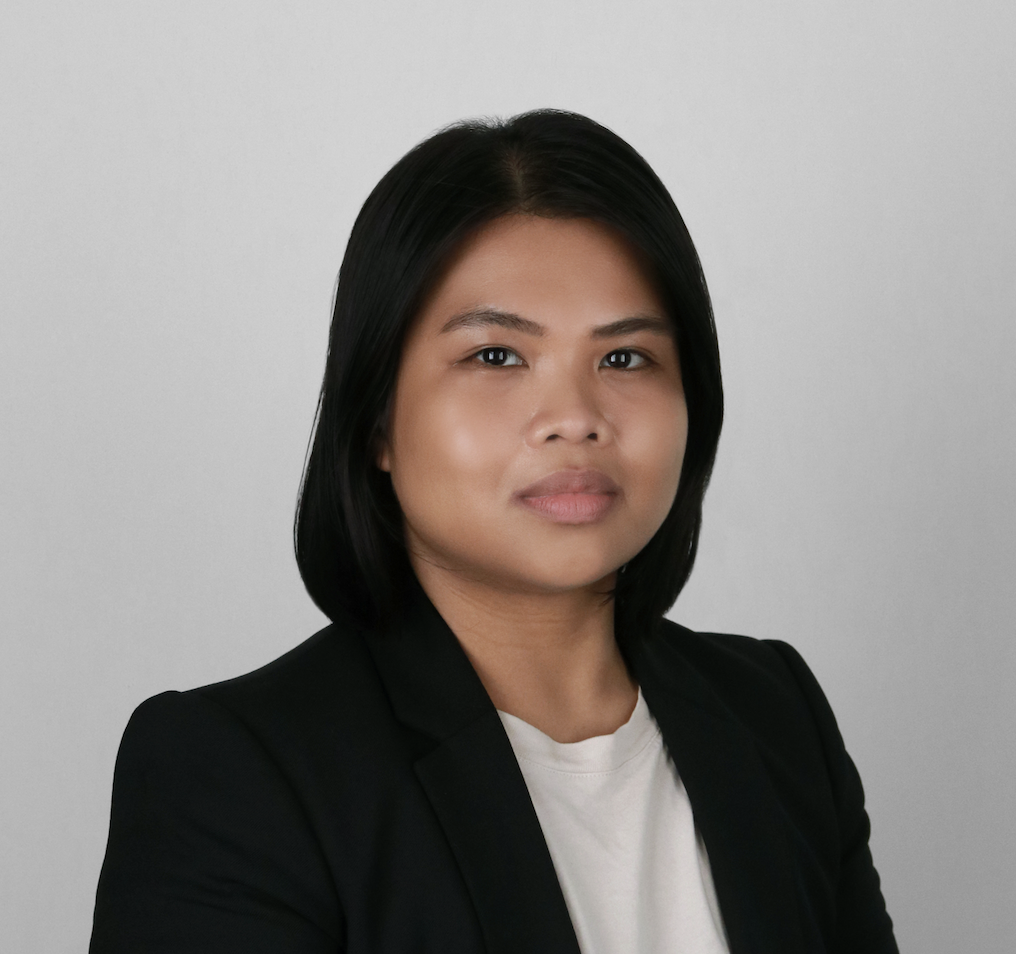}
Mary Lai O. Salva\~na is an Assistant Professor of Statistics at the University of Connecticut (UConn). Prior to joining UConn, she was a Postdoctoral Fellow at the Department of Mathematics at University of Houston. She received her Ph.D. in Statistics at the King Abdullah University of Science and Technology (KAUST), Saudi Arabia. She obtained her BS and MS degrees in Applied Mathematics from Ateneo de Manila University, Philippines, in 2015 and 2016, respectively. Her research interests include extreme and catastrophic events, risks, disasters, spatio-temporal statistics, environmental statistics, computational statistics, large-scale data science, and high-performance computing.
\endbio

\bio{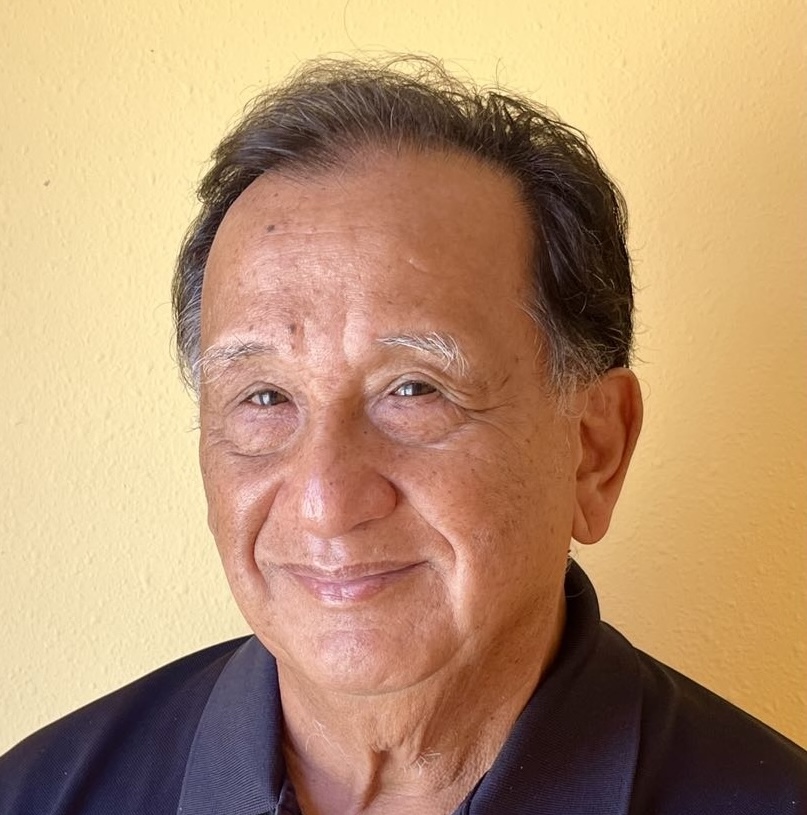}
Gregory L. Tangonan graduated in Physics from Ateneo de Manila University and earned his PhD in Applied Physics from the California Institute of Technology. He spent 32 years at Hughes Research Laboratories in Malibu, California, retiring as Lab Director of the Communications Laboratory. His expertise spans fiber optics, wireless communications, materials science, and complex systems science. He has published over 200 papers and holds 49 U.S. patents. After retiring, he returned to his alma mater and founded the Ateneo Innovation Center (AIC), which is actively engaged in disaster risk and resilience research. AIC has developed Resilience Hubs for nationwide deployment in the Philippines and has been at the forefront of research on cascaded disasters and risk assessment.
\endbio

\end{document}